\documentclass[letterpaper,twocolumn,10pt]{article}
\usepackage{usenix2019_v3}

\usepackage{tikz}
\usepackage{amsmath}
\usepackage{cite}
\usepackage{amsmath,amssymb,amsfonts}
\usepackage{bbm,epsfig,bm}
\usepackage{algorithmic}
\usepackage{graphicx}
\usepackage{textcomp}
\usepackage{xspace}
\usepackage[normalem]{ulem}
\usepackage[binary-units=true]{siunitx}
\usepackage{booktabs}
\def\BibTeX{{\rm B\kern-.05em{\sc i\kern-.025em b}\kern-.08em
    T\kern-.1667em\lower.7ex\hbox{E}\kern-.125emX}}
\usepackage[capitalise,noabbrev]{cleveref}
\usepackage{enumitem}

\usepackage{subcaption}
\usepackage{tcolorbox}
\usepackage{balance}
\usepackage{multirow}
\usepackage[numbers]{natbib}

\definecolor{dkgreen}{rgb}{0,0.5,0}
\definecolor{lessdkgreen}{rgb}{0,0.6,0}
\definecolor{dkred}{rgb}{0.5,0,0}
\definecolor{gray}{rgb}{0.5,0.5,0.5}

\usepackage{verbatimbox}
\usepackage{fancyvrb, listings}
\usepackage{courier}
\makeatletter
\AtBeginDocument{\DeclareCaptionSubType{lstlisting}
  \renewcommand{\p@sublstlisting}{\thelstlisting}
  
}

\newcommand{\mysec}[1]{\medskip \noindent \textbf{#1.}}

\hypersetup{
  hidelinks,
}
\makeatother
\newsavebox{\verbsavebox}

\newcommand{\eg}[0]{e.g., }
\newcommand{\ie}[0]{i.e., }

\newcommand{\GH}{{\sc GitHub}\xspace}
\newcommand{\osprey}{{\sc Osprey}\xspace}

\newcommand{\bv}[1]{\textcolor{purple}{\bf\small [#1 --BV]}}

\newcommand{\hide}[1]{}

\newcommand\lt[1]{{\lstinline[style=cstyle4]!#1!}}
\newcommand{\disappear}{\lt{<Component>}\xspace}
\renewcommand{\tt}{\texttt}

\newcommand{\modelname}{DIRTY\xspace}
\newcommand{\dataname}{DIRT\xspace}

\lstdefinestyle{cstyle}{
language=c,
basicstyle=\ttfamily\bfseries\scriptsize,
  morekeywords={virtualinvoke},
  keywordstyle=\color{blue},
  ndkeywordstyle=\color{red},
  commentstyle=\color{dkred},
  stringstyle=\color{dkgreen},
  numbers=left,
  breaklines=true,
  numberstyle=\ttfamily\footnotesize\color{gray},
  stepnumber=1,
  numbersep=10pt,
  backgroundcolor=\color{white},
  tabsize=4,
  showspaces=false,
  showstringspaces=false,
  xleftmargin=.23in,
  captionpos=b,
  escapeinside={$}{$},
  print
}

\lstdefinestyle{cstyle2}{
language=c,
basicstyle=\ttfamily\bfseries\scriptsize,
  morekeywords={virtualinvoke},
  keywordstyle=\color{blue},
  ndkeywords={a1,a2,a3,result},
  ndkeywordstyle=\color{red},
  commentstyle=\color{dkred},
  stringstyle=\color{dkgreen},
  numbers=none,
  breaklines=true,
  numbersep=10pt,
  backgroundcolor=\color{white},
  tabsize=2,
  showspaces=false,
  showstringspaces=false,
  xleftmargin=0,
  captionpos=b,
  escapeinside={$}{$},
  print
}

\lstdefinestyle{cstyle3}{
language=c,
basicstyle=\ttfamily\bfseries\scriptsize,
  morekeywords={virtualinvoke, Picture\_0, MpegEncContext\_0, AVCodecContext\_0},
  keywordstyle=\color{blue},
  ndkeywords={a1,a2,a3,v1,i,j},
  ndkeywordstyle=\color{red},
  commentstyle=\color{dkgreen},
  morecomment=[s]{<}{>},
  stringstyle=\color{dkgreen},
  numbers=none,
  breaklines=true,
  numbersep=10pt,
  backgroundcolor=\color{white},
  tabsize=2,
  showspaces=false,
  showstringspaces=false,
  xleftmargin=0,
  captionpos=b,
  escapeinside={$}{$},
  print
}

\lstdefinestyle{cstyle4}{
language=c,
basicstyle=\ttfamily\bfseries\scriptsize,
  morekeywords={virtualinvoke},
  keywordstyle=\color{blue},
  ndkeywords={result},
  ndkeywordstyle=\color{red},
  commentstyle=\color{dkred},
  stringstyle=\color{dkgreen},
  numbers=none,
  breaklines=true,
  numbersep=10pt,
  backgroundcolor=\color{white},
  tabsize=2,
  showspaces=false,
  showstringspaces=false,
  xleftmargin=0,
  captionpos=b,
  escapeinside={$}{$},
  print
}

\lstdefinestyle{cinlinestyle}{
language=c,
basicstyle=\ttfamily\bfseries\footnotesize,
  morekeywords={virtualinvoke, class, __int64, uint16_t, __int16, MpegEncContext_0},
  ndkeywords={a1,a2,a3,v1,i,j,c,ch,s,str,aN,vN,integer1,x,y,picture},
  keywordstyle=\color{blue},
  ndkeywordstyle=\color{red},
  commentstyle=\color{dkred},
  stringstyle=\color{dkgreen},
  escapeinside={$}{$},
  print
}

\lstdefinelanguage
    [hexrays]{Assembler}
    [x86masm]{Assembler}
    {morekeywords={addl,popq,pushq,movq,movl,retn,rbp,rsp},deletendkeywords={dword,ptr,short},morendkeywords={var1,var2}
    }

\lstdefinestyle{hexrays}{
  language=[hexrays]{Assembler},
  basicstyle=\ttfamily\bfseries\scriptsize,
  keywordstyle=\color{blue},
  ndkeywordstyle=\color{lessdkgreen},
  commentstyle=\color{dkred},
  print
}

\usepackage{float}

\usepackage{authblk}

\begin{document}

\date{}

\title{\Large \bf Augmenting Decompiler Output with Learned Variable Names and
  Types}

\author[*]{Qibin Chen}
\author[*]{Jeremy Lacomis}
\author[$\dagger$]{Edward J. Schwartz}
\author[*]{\authorcr Claire Le~Goues}
\author[*]{Graham Neubig}
\author[*]{Bogdan Vasilescu}
\affil[*]{Carnegie Mellon University. \{qibinc, jlacomis, clegoues, gneubig, bogdanv\}@cs.cmu.edu}
\affil[$\dagger$]{Carnegie Mellon University Software Engineering Institute. eschwartz@cert.org}
\maketitle

\begin{abstract}
A common tool used by security professionals for reverse-engineering binaries
found in the wild is the \emph{decompiler}. A decompiler attempts to reverse
compilation, transforming a binary to a higher-level language such as
C. High-level languages ease reasoning about programs by providing useful
abstractions such as loops, typed variables, and comments, but these
abstractions are lost during compilation. Decompilers are able to
deterministically reconstruct structural properties of code, but comments,
variable names, and custom variable types are technically impossible to recover.

In this paper we present DIRTY (DecompIled variable ReTYper), a novel
technique for improving the quality of decompiler output that automatically
generates meaningful variable names and types.
Empirical evaluation on a novel dataset of C code mined from GitHub shows
that DIRTY outperforms prior work approaches by a sizable margin,
recovering the original names written by developers 66.4\% of the time and
the original types 75.8\% of the time.
\end{abstract}

\section{Introduction}
Reverse engineering is an important problem in the context of software.
For example, security professionals use reverse engineering to understand the
behavior or provenance of malware~\cite{Durfina2013, Yakdan2015, Yakdan2016}, discover vulnerabilities in
libraries~\cite{VanEmmerik2007, Yakdan2016}, or patch bugs in legacy software~\cite{VanEmmerik2007, Yakdan2016}.
However, since it is rare to have access to source code, analysis is
often performed at the binary level.
This can be challenging: compilers optimize for execution speed or binary size,
not readability.

\looseness=-1
A number of tools attempt to make the process of examining
binary programs easier.
One is the \emph{disassembler}, which converts
raw binary code to a sequence of instructions executed by the compiler.
Although this produces human readable code, reasoning
about assembly code can still be difficult.
Operations that are simple to specify at a high-level are often translated into
a long sequence of assembly instructions (e.g., looping over the elements of an
array requires instructions that maintain an index variable, increment it each
iteration of a loop, and conditionally jump on its value).
Another, more sophisticated tool is a \emph{decompiler}, which
transforms code from binary to a high-level language such as C.

Although decompilers generate abstractions that improve code readability
and are widely used by reverse engineers in practice,
they never fully reconstruct the original developer-written code~\cite{Schulte2018},
since the process of compilation irrevocably destroys some information.
This means that useful pieces of information, such as comments, identifier names, and types,
all of which are known to meaningfully contribute to program comprehension~\cite{Gellenbeck1991, Lawrie2006},
are typically absent from decompiler output.
Nonetheless, recent work has shown that it \textit{is} possible to reconstruct some
useful information about the original code during decompilation, namely
identifier~\cite{Jaffe2018, lacomis2019dire} and procedural names~\cite{david2020neural},
\textit{even when this information is not part of the binary}.
The key insight is that human-written code is often repetitive in the same context
\cite{Hindle2012, Devanbu2015, allamanis2018survey}.
Therefore, given large corpora of human-written code, one can \emph{learn}
highly probable names for identifiers in similar contexts, even if not always
the exact names the authors of the code chose originally.
This is an important improvement on the facilities of modern decompilers, which
almost completely ignore names beyond simple heuristics
(e.g., \lt{i} and \lt{j} for loop guards).

In this paper we focus on the closely related problem of recovering meaningful variable
\emph{types}, an important
additional layer of code documentation that can help improve
readability and understandability~\cite{troshina2010reconstruction, Schulte2018, escalada2021improving}.
\cref{fig:example} shows an example of a simple function and its decompilation.
The author of the original code in \cref{fig:orig} has defined a \lt{pnt}
structure that contains two \lt{float} members used to refer to the X and Y
coordinates of a point.
This makes it possible to define a new point and refer to its members by name
(\eg \lt{p1.x} and \lt{p1.y}).
Because the decompiler does not know about the \lt{pnt} structure, it creates two
\lt{float} arrays instead of generating a \lt{struct} (\cref{fig:decomp}).
This can harm understandability. 
First, it is not clear that \lt{v1} and \lt{v2} represent points at all.
Second, even if better names were chosen, such as \lt{point1} and \lt{point2},
and a reverse engineer concluded that they represent 2D points, it is not
clear which array index refers to which coordinate, or even that the coordinates
are Cartesian (instead of \eg polar).

\begin{figure}
  \begin{lrbox}{\verbsavebox}
    \hspace{0.5cm}
    \begin{lstlisting}[style=cstyle2]
typedef struct point {
  float x;
  float y;
} pnt;

void fun() {
  pnt p1, p2;
  p1.x = 1.5;
  p1.y = 2.3;
  // ...
  use_pts(&p1, &p2);
}
    \end{lstlisting}
  \end{lrbox}
  \subcaptionbox{Original code\label{fig:orig}}{\usebox{\verbsavebox}}\hfill
  \begin{lrbox}{\verbsavebox}
      \begin{lstlisting}[style=cstyle2]
void fun() {
  float v1[2], v2[2];
  v1[0] = 1.5;
  v1[1] = 2.3;
  // ...
  use_pts(v1, v2);
}
      \end{lstlisting}
    \hspace{0.5cm}
  \end{lrbox}
  \subcaptionbox{Decompiled \lt{fun}\label{fig:decomp}}{\usebox{\verbsavebox}}\hfill
  \caption{A function with a \lt{struct} and its
    decompilation.\label{fig:example}}
\end{figure}

\begin{figure}
  \centering
    \begin{lrbox}{\verbsavebox}
    \hspace{0.6cm}
    \begin{lstlisting}[style=cstyle4]
void fun() {
  // stack layout:
  // [xxx][p][yyyy]
  char x[3];
  int y;
  // ...
}
\end{lstlisting}
\end{lrbox}
\subcaptionbox{Original code\label{fig:badorig}}{\usebox{\verbsavebox}}\hfill
    \begin{lrbox}{\verbsavebox}
      \begin{lstlisting}[style=cstyle4]
void fun() {
  // stack layout:
  // [xxxx][yyyy]
  char x[4];
  int y;
  // ...
}
      \end{lstlisting}
      \hspace{0.6cm}
      \end{lrbox}
      \subcaptionbox{Decompiled \lt{fun}\label{fig:baddecomp}}{\usebox{\verbsavebox}}\hfill
  \caption{A function illustrating the data layout problem in decompilation.
    In the stack layout the characters \lt{x}, \lt{y}, and \lt{p}
    represent a single byte assigned to the variables \lt{x} and \lt{y}, or
    padding data respectively. The decompiler cannot recognize that the inserted
    padding data does not belong to the \lt{x}
    array.\label{fig:baddata}}
\end{figure}

Unlike names, types are constrained by memory layouts, and thus theoretically
should be easier to recover (only types that fit that memory layout should be
considered as candidates).
In fact, decompilers already narrow down possible
type choices using the fact that base types targeting a specific platform can
only be assigned to variables with a specific memory layout (\eg on most
platforms an \lt{int} variable can never be retyped to a \lt{char} because they
require different amounts of memory).  This already makes it possible for
decompilers to infer base types and a small set of commonly-used
\lt{typedefs}.

On the other hand, despite performing a battery of complex binary analyses,
the data layout inferred by the decompiler is often incorrect, which makes
the problem harder.
For example, consider the program shown in \cref{fig:baddata}.
Two top-level variables are declared, \lt{x}: a
three-byte \lt{char} array, and \lt{y}: a four-byte \lt{int}.
During compilation, the compiler inserts a single byte of padding
after the \lt{x} array for alignment.
When this function is decompiled, the decompiler can tell where \lt{x} and
\lt{y} begin, but it cannot tell if \lt{x} is a three-byte array followed by a
single byte of padding, or a four-byte array whose last element is never used.

Prior work on reconstructing types falls into two groups.
The first, such as TIE~\cite{Lee2011}, attempt to recover \emph{syntactic} types, \eg \lt{struct \{float;
  float\}}, but not the names of the structure or its fields.
The second, such as REWARDS~\cite{lin2010automatic}, attempt to also recover the type name
(referred to as \emph{semantic} types).
However, these systems typically only support
a small set of manually-defined types and well-known library calls.
Neither the first nor the second deal with the padding issue above.

In contrast, our system \modelname (DecompIled variable ReTYper) recovers both
semantic and syntactic types, handles padding, and is not limited to a small set of manually-defined types.
Instead, \modelname supports 48,888 possible types encountered
``in the wild'' in open-source C code (compared to the 150 different type
names in 84 standard library calls supported by REWARDS). 
At a high level, \modelname is a \emph{Transformer-based~\cite{transformer} neural network
model to recommend types} in a particular context, which operates as a postprocessing
step to decompilation.
\modelname takes a decompiled function as input, and outputs probable
names and types for all of its variables.

To build \modelname, we start by mining open-source C code from \GH,
and then use a decompiler's typical ability to import variable names and types from
DWARF debugging information to create a \emph{parallel corpus of decompiled functions
with and without their corresponding original names and types}.
As a side effect of this large-scale mining effort, we also automatically compile a
\emph{library of types} encountered across our open-source corpus.
We then train \modelname on this data, introducing two task-specific innovations.
First, we use a Data Layout Encoder to incorporate memory layout information
into \modelname's predictions and simultaneously address a fundamental
limitation of decompilers caused by padding.
Second, we address both the variable renaming and retyping tasks simultaneously
with a joint Multi-Task architecture, enabling them to benefit from each other.

We show that \modelname{} can assign variable \textit{types} that agree with
those written by developers up to 75.8\% of the time, and \modelname also
outperforms prior work on variable \textit{names}.

Note that even though we implement \modelname on top of the Hex-Rays\footnote{\url{https://www.hex-rays.com/products/decompiler/}} decompiler
because of its positive reputation and its programmatic access to decompiler internals,
our approach is not fundamentally specific to Hex-Rays, and should conceptually work
with any decompiler that names variables using DWARF debug symbols.%

In summary, we contribute:
\begin{itemize}[itemsep=0.5pt,topsep=0pt]

\item \dataname---the Dataset for Idiomatic ReTyping---a large-scale
  public dataset of C code for training models to retype or rename decompiled code, consisting of
  nearly 1 million unique functions and 368 million code tokens.
\item \modelname---the DecompIler variable ReTYper---an open-source Transformer-based neural network model
  to recover syntactic and semantic types in decompiled variables.
  \modelname uses the \textit{data layout} of variables to improve retyping accuracy,
  and is able to \textit{simultaneously retype and rename} variables in decompiled code.
\end{itemize}
Example output from \modelname is available online at {\href{https://dirtdirty.github.io/explorer.html}{https://dirtdirty.github.io/explorer.html}}.

\section{Model Design}
In this section, we describe our machine learning model and decisions that
influenced its design, starting with some relevant background.
Our model is a \emph{neural network} with an \emph{encoder-decoder}
architecture.

\subsection{The Encoder-Decoder Architecture}
\label{bg:encoderdecoder}

Our task consists of generating variable types (and names) as
output given individual functions in decompiled code as input.
This means that unlike a traditional classification problem with a fixed number of classes,
both our input and output are sequences of variable length: input functions
(\eg fed into the network as a sequence of tokens) can have arbitrarily many variables,
each requiring a type (and name) prediction.

Therefore, we adopt an encoder-decoder architecture~\cite{Cho2014LearningPR},
commonly used for sequence-to-sequence transformations, 
as opposed to the traditional feed-forward neural network architecture used in classification
problems with a fixed-length input vector and prediction target.
More specifically, the \emph{encoder} takes the variable-length input and encodes it as a
fixed-length vector.
Then, this fixed-length encoding is passed to the \emph{decoder}, which converts
the fixed-length vector into a variable-length output sequence.
This architecture, further enhanced through the \emph{attention} mechanism~\cite{attention},
has been shown to be effective in many tasks such as machine translation, text
summarization~\cite{nallapati2016abstractive}, and image captioning~\cite{xu2015show}.

\subsection{Transformers}
\label{sec:transformer}

There are several ways to implement an encoder-decoder.
Until recently, the standard implementation used a particular type of recurrent neural network (RNN) with
specialized neurons called long short-term memory units; these neurons and
networks constructed from them are commonly referred to as
LSTMs~\cite{hochreiter1997long}.
More recently, Transformer-based models~\cite{raffel2020exploring,brown2020language,zaheer2020big,fedus2021switch},
building on the original Transformer architecture~\cite{transformer},
have been shown to outperform LSTMs and are considered to be the
state-of-the-art for a wide range of natural language processing tasks, including machine
translation~\cite{brown2020language}, question answering
and abstractive summarization~\cite{devlin2019bert, lewis2020bart},
and dialog systems~\cite{adiwardana2020towards}.
Transformer-based models have also been shown to outperform convolutional
neural networks such as ResNet~\cite{he2016deep} on image recognition tasks~\cite{dosovitskiy2020image}.

Transformers have several properties that make them a particularly good fit for
our type prediction task.
First, they capture long-range dependencies,
  which commonly occur in program code,  more effectively than RNNs.  For example, a variable declared at the
  beginning of a function may not be used until much later; an ideal model
  captures information about all uses of a variable.
Second, transformers can perform more computations in parallel on typical GPUs
  than LSTMs.  As a result, training is faster, and a Transformer can train on
  more data in the same amount of time.  In our case, this enables us to train
  on our large-scale, real-world dataset, which consists of 368 \emph{million}
  decompiled code tokens.

Although there have been a number of advances in neural machine translation
since the original Transformer model~\cite{transformer}, most recent advances
focus on improvements on other factors, such as
training data and objectives~\cite{devlin2019bert,lewis2020bart,raffel2020exploring,brown2020language},
dealing with longer sequences~\cite{zaheer2020big},
efficiency~\cite{child2019generating},
and scaling~\cite{fedus2021switch},
rather than changing the fundamental architecture.
Moreover, most of these improvements are tailored for
the natural language domain, making them less generalizable than the
original model and inapplicable to our task.  Instead, we keep
our model simple, which allows different, better architectures
or implementations to be used out-of-the-box in the future.  For example,
the recent Vision Transformer (ViT)~\cite{dosovitskiy2020image}, which
also intentionally follows the original Transformer architecture ``as closely as possible''
when adapting Transformers to computer vision tasks.

We omit the technical details of Transformers, including multi-headed
self-attention, positional encoding, and the
specifics of training as they are beyond the scope of this paper.

\subsection{DIRTY's Architecture}
\label{sec:dirty_arch}
\begin{figure*}
  \centering
  \includegraphics[width=0.9\textwidth]{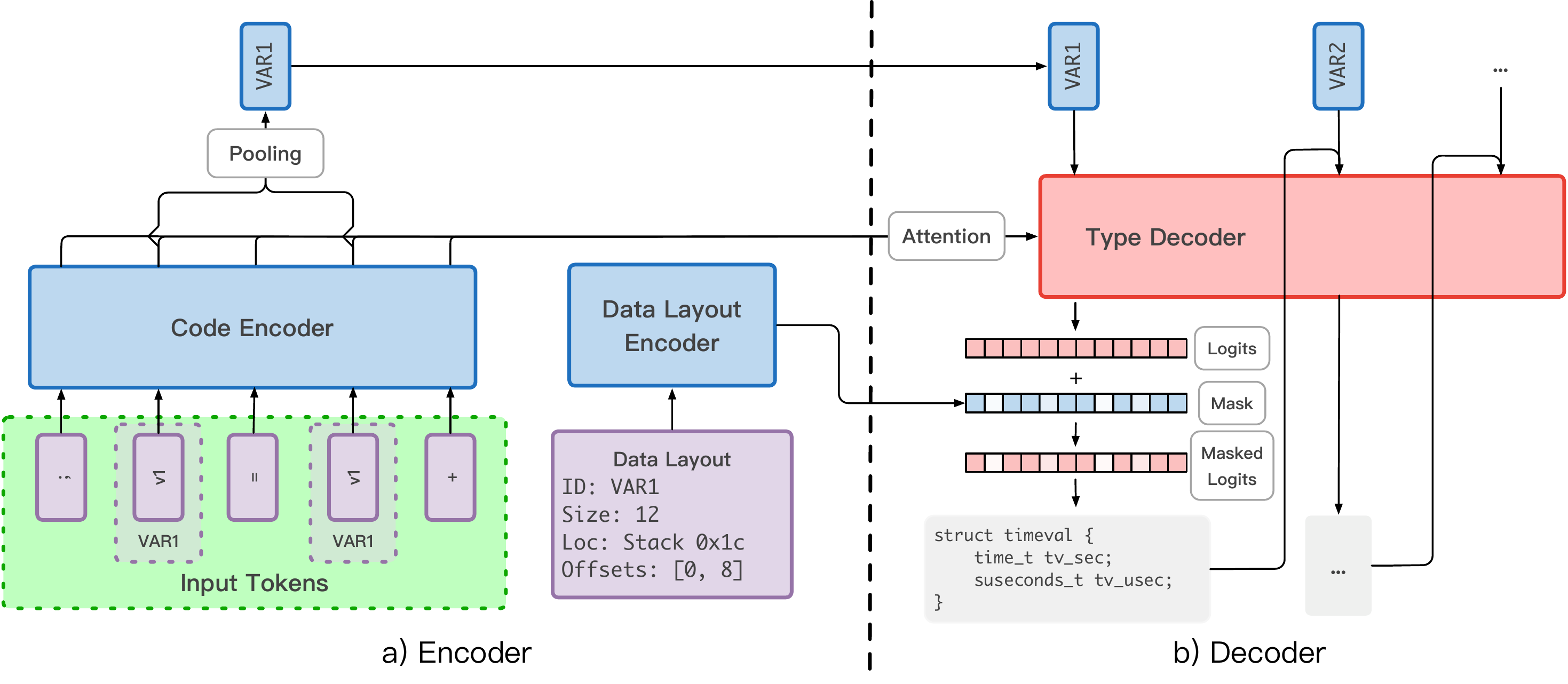}
  \caption{\label{fig:model}Overview of \modelname{}'s neural model architecture
    for predicting types. Decompiled code is sequentially fed into
    the Code Encoder. When the input of the code encoder corresponds to a
    specific variable (\eg \lt{VAR1}), it is pooled with other instances of the
    same variable to generate a single encoding for that variable. Each pooled
    encoding is then passed into the Type Decoder, which outputs a vector of the
    log-odds (logits) for predicted types. This vector is masked with a vector
    generated by the Data Layout encoder and the most probable type is chosen
    from the masked logits.}
\end{figure*}

In \modelname, we cast the retyping problem as a transformation from a
sequence of tokens representing the decompiled code to a sequence of
types, one for each variable in the original source code.
This section describes \modelname's architecture in detail.
\cref{fig:model} shows an overview of the architecture.

\setlength{\abovedisplayskip}{2pt}
\setlength{\belowdisplayskip}{2pt}
\setlength{\abovedisplayshortskip}{0pt}
\setlength{\belowdisplayshortskip}{0pt}

\mysec{Code Encoder}
The encoder converts the sequence of code tokens of
the decompiled function (lower-left of \cref{fig:model}),
$x=\left(x_{1}, x_{2}, \ldots, x_{n}\right)$,
into a sequence of representations,
\begin{align}
\mathbf{H} = \left(\mathbf{h_1}, \mathbf{h_2}, \ldots, \mathbf{h_n}
  \right),
\end{align}
where each continuous vector $\mathbf{h_i}\in \mathbb{R}^{d\_model}$ is the \textit{contextualized representation} for the $i$-th token $x_i$.
During training, the encoder learns to encode the information in the decompiled
function $x$ relevant to solving the task into $\mathbf{H}$.
For example, for a code token $x_i = $\lt{v1}, useful information about \lt{v1}
in the context of $x$ (\eg operations performed on \lt{v1}) is
automatically learned and stored in $\mathbf{h_i}$.

Specifically, we denote the encoding procedure as
\begin{align}
  \mathbf{H} = f_{en}\left(x;\theta_{en}\right),
\end{align}
\noindent
where the input $x=\left(x_{1}, x_{2}, \ldots, x_{n}\right)$ is the
code token sequence of the decompiled function and the output
$\mathbf{H} = \left(\mathbf{h_1}, \mathbf{h_2}, \ldots, \mathbf{h_n}
\right)$ is the sequence of deep contextualized representations.
$f_{en}$~denotes the encoder, implemented with neural networks,
and $\theta_{en}$ denotes its learnable parameters.

The ultimate goal of \modelname{} is to make type predictions about
each \emph{variable} that appears in the decompiled function.
However, the encoder produces hidden representations for every \emph{code token}
(\eg{``\lt{v1}'', ``\lt{:}'', ``\lt{=}'', ``\lt{v1}'', ``\lt{+}'', ``\lt{1}'' are all tokens}).
Because a variable can appear multiple times in the code tokens of a function,
we need a way to summarize all appearances of a variable.
We achieve this through \emph{pooling}, where the representation for the $t$-th
variable\footnote{$t$ is commonly used in RNN literature because it refers to a
  ``timestep''.} is computed based on all of its appearances in the code tokens,
$A_t,$ using an average pooling operation~\cite{lacomis2019dire}
\begin{align}
  \mathbf{v_t} = \operatorname{AveragePool}_{x_i \in A_t}\mathbf{h_i},\ t=1,\ldots, m
\end{align}
where $m$ is the number of variables in the function.
This solution removes the burden of gathering all information about a
variable throughout the function into a single token representation from the
model.
The pooled representation for the first variable, \texttt{VAR1},
is shown in the upper-left of \cref{fig:model}.

\mysec{Type Decoder}
Given the encoding of the decompiled tokens, the decoder predicts the
most probable (\ie idiomatic) types for all variables in the function.
The decoder takes the encoded representations of the code tokens ($\mathbf{H}$)
and identifiers ($\mathbf{v_t}$) as input and
predicts the original types $\hat{y} = (\hat{y}_1, \hat{y}_2, \ldots,
\hat{y}_m)$ for all $m$ variables in the function.
Unlike the encoder, the decoder predicts the output step-by-step using former
predictions as input for later ones.\footnote{This is known as an
  \emph{autoregressive} model.}

At each time step $t$, the decoder tries to predict the type for the $t$-th variable as follows:
\begin{enumerate}[itemsep=0pt,topsep=1pt]
  \item The decoder takes the code representations $\mathbf{H}$ and variable
  representation $\mathbf{v_t}$ from the encoder, and also previous predictions
  $\hat{y}_1, \hat{y}_2,\ldots,\hat{y}_{t-1}$ from itself, to compute a
  hidden representation $\mathbf{z_t}\in\mathbb{R}^{d\_model}$
  \begin{align}
    \mathbf{z_t} = f_{de}(\hat{y}_1, \hat{y}_2,\ldots,\hat{y}_{t-1}, \mathbf{v_t},\mathbf{H};\theta_{de})
  \end{align}
  where $f_{de}$, $\theta_{de}$ denotes the decoder and its parameters.
  The hidden representation $\mathbf{z_t}$ is then used for prediction.
  \item The output layer of the decoder
  then uses its learnable weight matrix $\mathbf{W}$
  and bias vector $\mathbf{b}$ to transform the
  hidden representation $\mathbf{z_t}$ to the \textit{logits} for prediction
  \begin{align}
    \mathbf{s_t} = \mathbf{W}\mathbf{z_t}+\mathbf{b}, \label{eq:output_layer}
  \end{align}
  where $\mathbf{s_t}\in \mathbb{R}^{|\mathcal{T}|}$,
  $\mathbf{W} \in \mathbb{R}^{|\mathcal{T}| \times d\_model}$,
  $\mathbf{b} \in \mathbb{R}^{|\mathcal{T}|}$,
  and $|\mathcal{T}|$ is the number of types in the type library.
  The logits $\mathbf{s_t}$ is the unnormalized probability predicted
  by the model, or the model's \textit{scores} on all types
  \item Finally, the $\operatorname{softmax}$ function computes a probability distribution
  over all possible types from $\mathbf{s_t}$
  \begin{align}
    \operatorname{Pr}(\hat{y}_t|\hat{y}_1, \hat{y}_2,\ldots,\hat{y}_{t-1}, x) = \operatorname{softmax}\mathbf{s_t}
  \end{align}
\end{enumerate}

\noindent

Note that the type library $\mathcal{T}$ is fixed, meaning
\modelname can only predict types that it has seen during training. We discuss
this limitation, its implications, and potential mitigations in
\cref{sec:future}. However, \modelname can recover structure types as well as normal
types, as both are simply entries in $\mathcal{T}$.

The goal of the decoder is to find the \emph{optimal}
set of type predictions for all variables in a given function
(\ie the predictions with the highest combined probability):
$\operatorname{argmax}_{\hat{y}}\operatorname{Pr}(\hat{y}|x).$
This probability can be factorized as the product of probabilities at each step:
\begin{align}
  \operatorname{Pr}(\hat{y} \mid x) =
    \prod_{t=1}^{m}
    \operatorname{Pr}\left(
      \hat{y}_{t} \mid \hat{y}_{1}, \hat{y}_{2}, \ldots \hat{y}_{t-1}, x
    \right).
  \label{eq:prob_factor}
\end{align}
We've shown how to compute
$\operatorname{Pr}\left(\hat{y}_{t} \mid \hat{y}_{1}, \hat{y}_{2}, \ldots \hat{y}_{t-1}, x\right)$
with the decoder, but finding the optimal $\hat{y}=\left(\hat{y}_{1}, \hat{y}_{2}, \ldots, \hat{y}_{m}\right)$
is not an easy task, because
each variable can have $|\mathcal{T}|$ possible predictions, and each
prediction affects subsequent predictions. The time complexity of exhaustive search is
$\mathcal{O}(|\mathcal{T}|^{m})$.
Therefore, finding the optimal
prediction is often computationally infeasible for large functions.
A simple approach is \emph{greedy decoding},
selecting the most promising prediction at
every step based on the previously selected predictions,
  i.e., taking the max
$\hat{y}_t = \operatorname{argmax}_{y_t}Pr\left(\hat{y}_{t}=i \mid
  \hat{y}_{1}, \hat{y}_{2}, \ldots, \hat{y}_{t-1}, x\right)$.
Greedy decoding is fast, but it often finds subpar predictions.

In \modelname, we use beam search~\cite{ney1987data}, a compromise between
greedy decoding and an exhaustive search.
Rather than only taking the most promising prediction (greedy),
beam search considers a configurable number of most
promising predictions at each step.  In practice, it is usually able
to find good (but not optimal) predictions, but is significantly
faster than an exhaustive search.

\subsection{Data Layout Encoder}
\label{sec:datalayout}

The model described so far only uses information encoded into
the code tokens of the decompiled representation.
But to actually \emph{create} such an output, decompilers typically
perform a battery of complex binary analyses.  Some decompilers
allow the user to programmatically access the interim
results from some of these analyses.
In particular, Hex-Rays provides information about
the storage location (\eg register or stack offset),
size, nested data types (\eg if the variable is a \lt{struct}), and offsets of
its members, if any (\eg offsets in an array or of fields in a \lt{struct}),
for each variable in a function.
Intuitively, this information can help \modelname rule out bad
predictions.  For instance, a variable that is 4 bytes long could not
be a \lt{char} type because it would not fit.

One inefficient approach could use this information as a hard
constraint on the decoder's predictions, \ie a \emph{mask} which sets the
probability of any ``incompatible'' types to 0.
However, this runs into a problem when the decompiler incorrectly reconstructs the
data layout (see \cref{fig:baddata}).
To mitigate this, \modelname learns a \emph{soft} mask, reducing probabilities
without setting them to 0.
For example, \modelname can learn based on many observations that a decompiled
\lt{char[4]} should be typed as a \lt{char[3]} 5\% of the time and \lt{char[4]}
80\%, and adjust the predictions of the type decoder accordingly.
This allows the model to learn how best to incorporate the data layout
information from the decompiler, including when the information is
likely to be incorrect. \cref{fig:model} illustrates where
the data layout encoder fits into the overall architecture.

\begin{figure}
  \centering
  \includegraphics[width=0.8\linewidth]{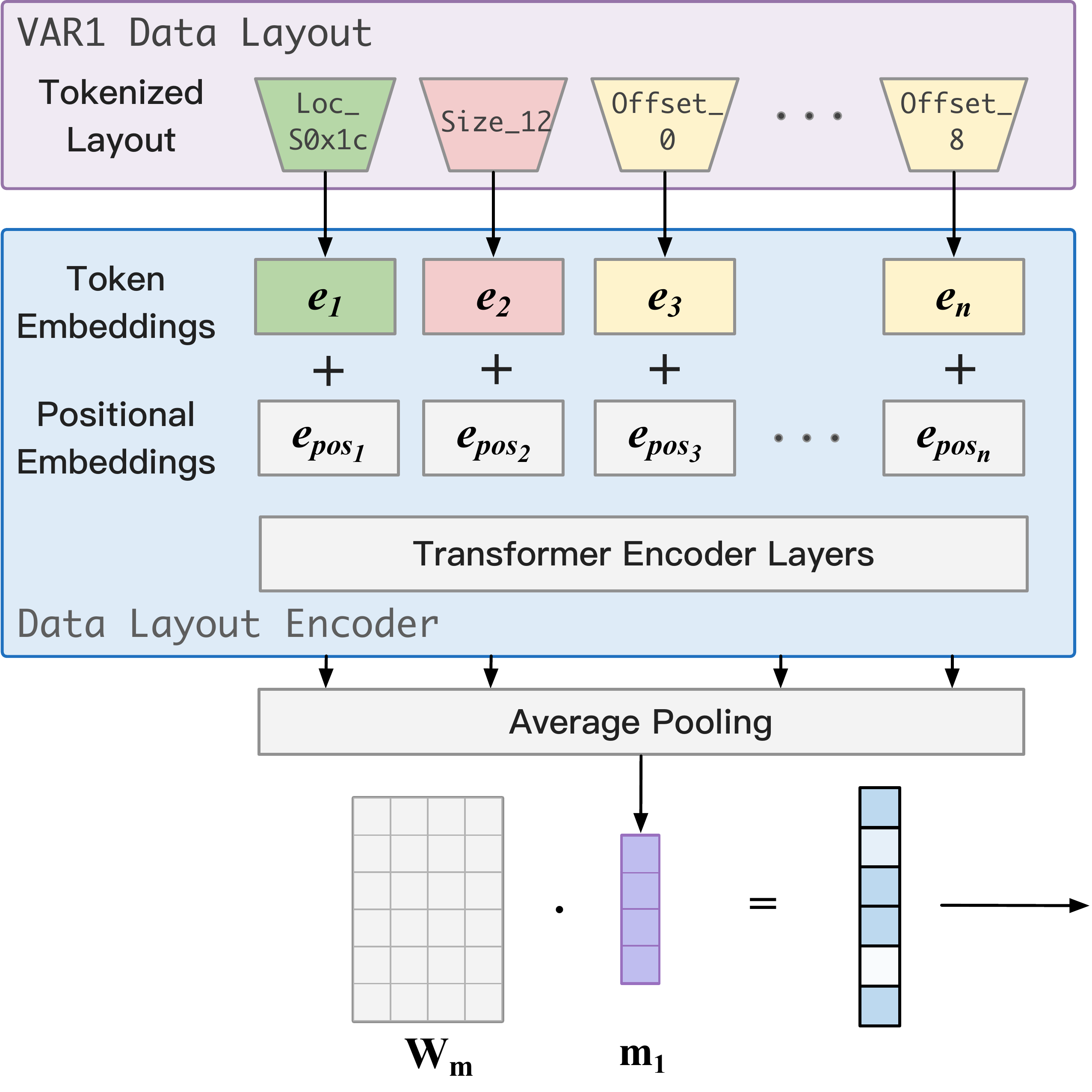}
  \caption{The Data Layout Encoder of \modelname{}. The data layout for a
    specific variable, including its location, size, and offsets of its members
    is passed into the layout encoder (top), generating a mask (bottom).}
  \label{fig:mem}
\end{figure}

To implement the soft mask encoder, we jointly train another Transformer
encoder to use data layout information to generate a mask.
\cref{fig:mem} shows the internals of the data layout encoder.
First, variable data layout is passed to the encoder.
There are three parts to the data layout for a specific variable, each of which
is simply converted to a token:
\begin{description}[itemsep=0pt,topsep=1pt]
\item[Location:] A variable can be located either 
  in registers (tokenized as \tt{[Loc\_<Register Name>]}) or on the stack (\tt{[Loc\_S<Offset>]}).
	E.g., a variable stored 28 bytes below the stack pointer is tokenized as \tt{[Loc\_S0x1c]}.
  \item[Size:] Measured in bytes and tokenized as
    \tt{[Size\_<Size>]}.
  \item[Internal Offsets:] The offsets of members of the type (either array elements or struct fields), in bytes.
    E.g., the type \lt{int[2]} would have the offsets
    $\{0, 4\}$, while a \lt{struct} with two \lt{char} fields would have the
    offsets $\{0, 1\}$.
    These are tokenized as a sequence of \tt{[Offset\_<Offset>]}.
    For consistency, we also use \tt{[Offset\_0]} for types without substructure
    (\ie scalar types like \lt{int}).
\end{description}
The tokenized data layout information is concatenated into a sequence denoted $M_t$
and then encoded as
\begin{align}
  \mathbf{m}_t = f_{layout}\left(M_t;\theta_{layout}\right),
\end{align}
where $\mathbf{m}_t$ is the hidden representation of data layout
information.  Inspired by~\citet{michel2018extreme},
we adjust the output type distribution with data layout information.  Formally, we modify
\cref{eq:output_layer} to fuse the data layout representation
$\mathbf{m}_t$ into the final output layer:
\begin{align}
  \tilde{\mathbf{s}}_{t} = \mathbf{s}_t + \mathbf{W}_m\mathbf{m}_t = \mathbf{W}\mathbf{z}_{t} + \mathbf{W}_m\mathbf{m}_t + \mathbf{b},
  \label{eq:mem}
\end{align}
where $\mathbf{s}_t$ is the logits predicted by the Type Decoder,
$\mathbf{W}_m\mathbf{m}_t$ is the ``soft mask'' produced by
the Data Layout Encoder,
and $\tilde{\mathbf{s}}_t$ is the new masked logits.
$\mathbf{W}_m \in \mathbb{R}^{|\mathcal{T}|\times d\_model}$ denotes the learnable
weight matrix in the final layer of Data Layout Encoder for transforming the
data layout representation $\mathbf{m}_t \in \mathbb{R}^{d\_model}$ to the mask
$\in \mathbb{R}^{|\mathcal{T}|}$. This implements a soft filter
for type prediction using data layout information.

\subsection{Multi-Task}
\label{sec:multitask}

Many variable names are indicative of their underlying type.
For example, \lt{i} and \lt{j} are often used to represent integers,
\lt{s} and \lt{str} are often used to represent strings, etc.
Thus, intuitively, there is some connection between a variable's
\emph{name} and its \emph{type}.
Indeed, measuring the adjusted mutual
information~\cite{vinh2010information} between variable names and types in our dataset,
we find a moderate association ($0.41$ on the scale $[0,1]$).
Since variable \emph{names} can often be recovered
from decompiled code using neural models~\cite{lacomis2019dire},
this may help us learn to predict variable types as well (and vice versa).

\begin{figure}
  \centering
  \includegraphics[width=0.8\linewidth]{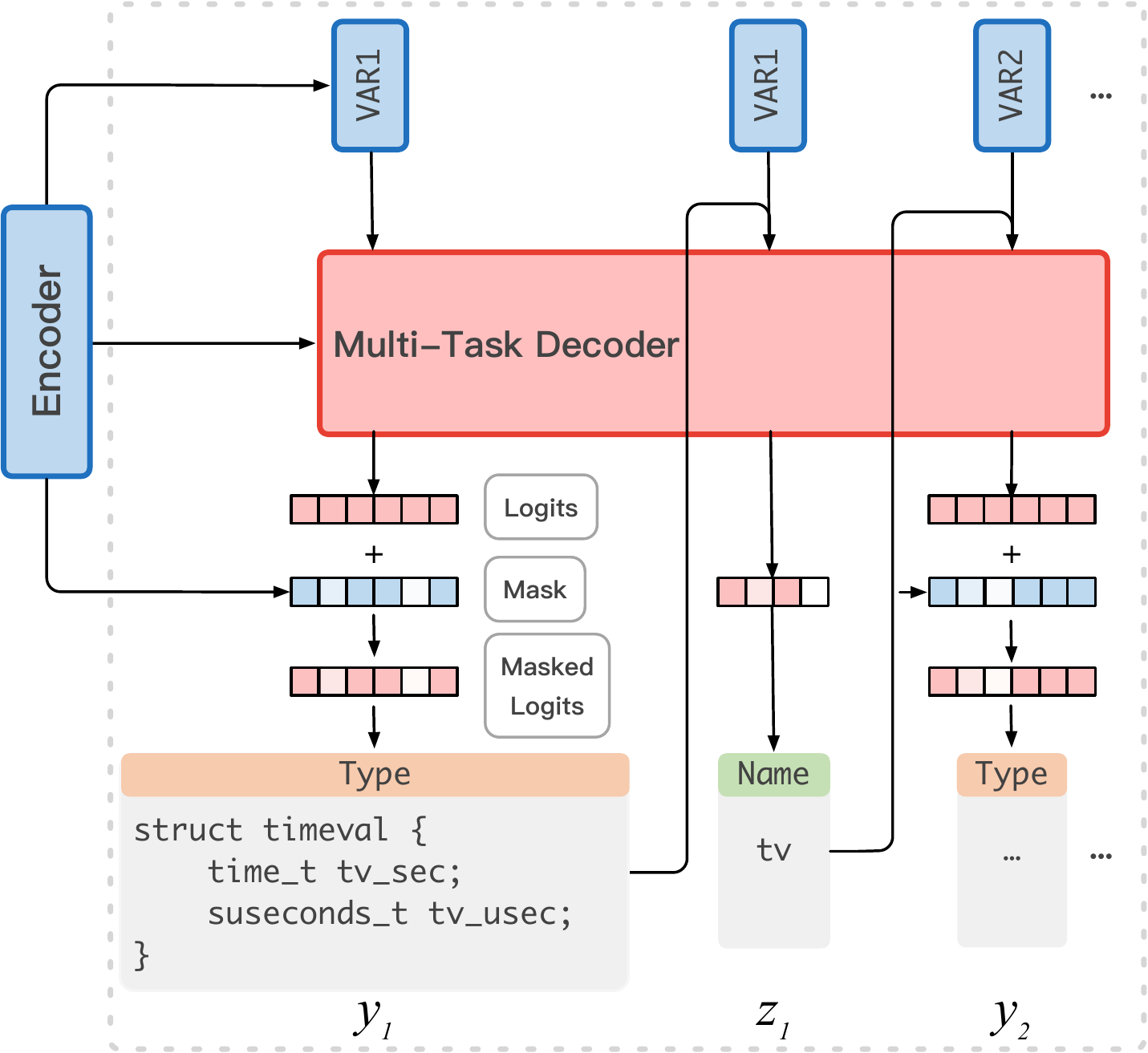}
  \caption{The multi-task decoder for \modelname{}, which predicts both variable
    types and names. The encoder architecture is the same as in
    \cref{fig:model}. Each variable is passed to the decoder twice, the first
    time a type is predicted ($y_i$), and the second time a name is
    predicted ($z_i$). Note that the data layout encoding of a variable is only used
    to weight type predictions.}
  \label{fig:mt}
\end{figure}

To test this, we extend \modelname to also predict names with a single,
integrated multi-task model. That is, we also predict a variable name for each
variable in the function
\begin{align}
  \hat{z}=\left(\hat{z}_{1}, \hat{z}_{2}, \ldots, \hat{z}_{m}\right)
\end{align}
where $\hat{z}_t$ denotes the predicted name for the $t$-th variable.

\modelname's decoder outputs are interleaved to predict names and
types in parallel (\cref{fig:mt}).
The first time the decoder is invoked on the $t$-th variable, it outputs the
predicted \emph{type} ($\hat{y}_t$) and the second time it outputs a predicted
\emph{name} ($\hat{z}_t$).

The training and prediction procedures remain almost the same, with
two notable exceptions.
First, to improve performance, the Data Layout encoder is
\emph{not} activated when the decoder is predicting a variable's name.
This is unnecessary because name prediction depends on the predicted type, which
has already incorporated the data layout information.
Preliminary experiments confirmed no improvement in accuracy when
using the Data Layout encoder for name prediction.

Second, there are two ways to interleave the predictions of types and names:
types first or names first.
In theory, this does not matter because they are equivalent if the learned model
and the decoding algorithm are ideal.
In practice, we chose to predict types first because we believe the type
prediction task should be easier (since there is more information) and it better
reflects how developers define variables.

\section{Evaluation}

We conducted experiments to evaluate \modelname, answering the following
research questions:

\begin{description}[itemsep=0pt,topsep=1pt]
  \item[RQ1:] How effective is \modelname{} at idiomatic retyping?
  \item[RQ2:] How well does \modelname{} perform on other decompilation benchmarks compared to prior work?
  \item[RQ3:] How does each component of \modelname{} contribute to the retyping and renaming performance?
  \item[RQ4:] How does compiler optimization affect \modelname{}'s prediction accuracy?
\end{description}

\subsection{Experimental Setup}
\label{sec:setup}
First, we introduce the \dataname dataset we used for training \modelname, and
experimental setup details.
The detailed hyperparameters for our deep learning model and environment
configuration are described \cref{sec:appendix}.

\mysec{Dataset for Idiomatic ReTyping (DIRT)}
To create \dataname, we queried a 2017 version of the
\textsc{GHTorrent}\footnote{\url{https://ghtorrent.org}} database, compiling a list
of public \GH repositories predominantly written in C.
We then cloned these repositories locally using an open-source tool, GHCC,\footnote{\url{https://github.com/huzecong/ghcc}}
to automatically build them.
GHCC identifies build instructions (e.g., Makefiles) in repositories, creates a
Docker container with the requisite libraries, and attempts to build the
project.  We used GCC version 9.2.0.  For most experiments, we explicitly
disable optimizations using the \tt{-O0} compiler flag.  We also evaluated
\modelname at higher optimization levels in \cref{sec:optimization_levels}.
This process resulted in 4,346,134 automatically compiled 64-bit
x86 binaries.
After compilation, we then decompiled each binary using Hex-Rays and
filtered out any functions that did not have variables requiring
renaming or retyping. Following
DIRE~\cite{lacomis2019dire}, we compiled each binary again
with debugging information to align decompiler-assigned
variable names (e.g., \lt{v1}) and developer-assigned variable names
(e.g., \lt{picture}) to form training examples.

Since DIRE was only concerned with renaming, its dataset did not include
variables which did not correspond to a named variable in the
original source code.
Many such variables are actually caused by mistakes in the decompiler
during type recovery, for instance decompiling a structure to multiple
scalar variables instead.
Since the goal of \dataname is to enable type recovery and fix such
mistakes, we label these instances as \disappear to denote that
they are \emph{components} of a variable in the source code.  This
allows the model to combine them with other variables into an array
or a struct.

The final DIRT dataset consists of 75,656 binaries randomly sampled
from the full set of 4,346,134 binaries to yield a dataset that we
could fully process based on the computational resources we had
available.
We split the dataset per-binary
as opposed to per-function, which
ensures that different functions from the same binary cannot be in
both the test and training sets.
The training
dataset consists of 997,632 decompiled functions, and a total number
of 48,888 different types.  We also preprocess the decompiled code
with byte-pair encoding (BPE)~\cite{sennrich2015neural}, a widely
adopted technique in NLP tasks to represent rare words with limited
vocabulary by tokenizing them into subword units.  After this step,
the \dataname{} dataset consists of 368 \emph{million} decompiled code
tokens, and an average of 220.3 tokens per function.
Detailed statistics about the \dataname dataset and the train/valid/test split
can be found in \cref{tab:dataset} in \cref{sec:appendix}.

\mysec{Metrics} We evaluate \modelname using two metrics:

\begin{description}[itemsep=0pt,topsep=1pt]
  \item[\textit{Name Match}:]
Following DIRE~\cite{lacomis2019dire}, we consider a variable name prediction
correct if it exactly string matches the name assigned by the original
developer.  We compute the prediction accuracy as the average percentage of
correct predictions across all functions in the test set.

  \item[\textit{Type Match}:]
We consider a type prediction to be
      correct only if the predicted type fully matches the ground
      truth type, including data layout, and the type and name of any
      fields if applicable.
      We serialize types to strings and use string matching to determine type
      matching.
\end{description}

Note that both metrics are conservative. Predictions may still be meaningful, even if
not identical to the original names. A human study evaluating the quality of predicted
types and names is beyond the scope of the current paper.

\mysec{Meaningful Subsets of the Test Data}
We introduce several subsets of the \dataname{} test set to better
interpret the results:
\begin{description}[itemsep=0pt,topsep=1pt]
\item[\textit{Function in training} vs \textit{Function not in
    training.}] Similarly to \citet{lacomis2019dire}, \textit{Function in training} consists
  of the functions in the test set that also appear in the training
  set,
  which are mainly library functions.
  Allowing this duplication simulates the realistic use case of
  analyzing a new binary that uses well-known libraries.
  We also
  separately measure the cases where the function is not known during
  training (\ie \textit{Function not in training}) to measure the
  model's generalizability.
\item[\textit{Structure types}.] Only 1.8\% of variables in \dataname have
  structure types.  Because of this low percentage, examining overall
  accuracy may not reflect \modelname's accuracy when predicting
  structure types, which we have found anecdotally to be more
  challenging.  To mitigate this, we separately measure \modelname's
  accuracy on structures in addition to its overall accuracy.
\end{description}

\subsection{RQ1: Overall Effectiveness}

We evaluate \modelname on the idiomatic retyping task and report its
accuracy compared to several baselines.

\mysec{Baselines}
We measure our accuracy with respect to two baseline methods for predicting variable types:
\begin{description}[itemsep=0pt,topsep=1pt]
\item[\textit{Frequency by Size}] The number of bytes a variable occupies is the most basic information for a type.
  For this technique, we predict the most common developer-assigned type for a
  given size (as reported by the decompiler).
  E.g., \lt{int} is the most common 4-byte type, and \lt{\_\_int64} is
  the most common 8-byte type; this baseline simply assigns these types to
  variables of the respective size.
\item[\textit{Hex-Rays~\cite{hexrays}}]
  During decompilation, Hex-Rays already predicts a type for
  each variable, so we can use these predictions as a baseline.
  However, Hex-Rays cannot
  predict developer-generated types without prior knowledge of them, \eg
  Hex-Rays assigns \lt{unsigned \_\_int16}
  instead of the more common \lt{uint16\_t}, which puts it at an
  unfair disadvantage.  For this baseline, we reassign the type chosen
  by Hex-Rays to the most common developer-chosen name associated with
  it (\eg we replace every \lt{unsigned \_\_int16} with
  \lt{uint16\_t}.

\end{description}

\mysec{Results}
\label{sec:main_results}
As shown in \cref{tab:accuracy}, \modelname{}
can correctly recover 75.8\% of the original (developer-written) types
from the decompiled code.  In contrast, Hex-Rays, the highest scoring
baseline, can only recover 37.9\% of the original types.

\begin{table}
    \centering
    \begin{tabular}{lrrrrrr}
      \toprule
      & \multicolumn{2}{c}{Overall} & \multicolumn{2}{c}{In Train} & \multicolumn{2}{c}{Not in Train} \\
      Method & All & Struct & All & Struct & All & Struct  \\
      \midrule
      F$_{\text{Size}}$ & 23.6 & 9.7  & 23.5 & 9.1  & 23.8 & 10.4 \\
      HR               & 37.9 & 28.7 & 39.0 & 28.7 & 36.4 & 28.7 \\
      \modelname{}     & \textbf{75.8} & \textbf{68.6} & \textbf{89.9} & \textbf{79.2} & \textbf{56.4} & \textbf{54.6} \\
      \bottomrule
    \end{tabular}
    \caption{\modelname has higher retyping accuracy than
    Frequency By Size (F$_{\text{Size}}$) and Hex-Rays (HR)
    on the \dataname{} dataset, both for
      all types (All) and on structural types alone (Struct).}
    \label{tab:accuracy}
\end{table}

As expected, \modelname performs even better when it has seen a
particular function before (In Train), generating the same type as the developer
89.9\% of the time.  This indicates that \modelname{} works
particularly well on common code such as libraries.  Even when a
function has never been seen (Not in Train), \modelname{} predicts the correct type
56.4\% of the time.

\cref{tab:accuracy} also shows the performance of \modelname{} on
structure types alone.  Correctly predicting structure types is more
difficult than predicting scalar types, and all models show a drop in
performance.  Despite this drop, \modelname{} still achieves 68.6\%
accuracy overall, and 54.6\% accuracy on
the \emph{Function not in training} category.
Frequency By Size struggles on structures with only 9.7\%
accuracy; this is expected since structures of a given size
can have many possible types.  Hex-Rays is slightly more accurate at
28.7\%, as the decompiler is able to analyze the layout of structures.

\begin{table*}
    \centering
    \small
    \begin{tabular}{lrlrlrlr}
        \toprule
        \multicolumn{2}{c}{\lt{int}} & \multicolumn{2}{c}{\lt{char *}} & \multicolumn{2}{c}{\lt{class std\:\:string}} \\
        \midrule
        \lt{int}          & 88.8\%   & \lt{char *}       & 60.3\%      & \lt{class std\:\:string}                 & 47.5\% \\
        \lt{unsigned int} &  4.3\%   & \lt{const char *} & 11.4\%      & \lt{char[32]}                             & 24.2\% \\
        \disappear        &  2.7\%   & \disappear        &  4.4\%      & \lt{char[47]}                             & 14.6\% \\
        \lt{uint32\_t}    &  0.8\%   & \lt{\_\_int64}    &  4.1\%      & \lt{class std::\_\_cxx11::basic\_string} &  6.1\% \\
        \lt{u\_int32\_t}  &  0.3\%   & \lt{size\_t}      &  1.8\%      & \lt{char[40]}                             &  3.5\% \\
        \bottomrule
    \end{tabular}
    \caption{Example variable types from the \textit{Function not in
        training} testing partition. The top rows are the
      developer-assigned types and the columns show \modelname{}’s
      top-5 most frequent predictions. \disappear represents a
      prediction that the variable in the decompiled code does not
      correspond to a variable in the source code (\eg because it
      corresponds to a member of a struct).  }
    \label{tab:preds}
\end{table*}

\cref{tab:preds} shows several examples of retyping predictions from
the \textit{Function not in training} partition.
These examples show that accuracy is not the full story; even when
\modelname{} is unable to predict the correct type, the differences
are often minor (\eg \lt{unsigned int} v.{} \lt{int}, and \lt{const char
  *} v.{} \lt{char *}).
The bottom half of \cref{tab:preds} shows prediction examples of
structure types.\footnote{We omit the full predicted contents of
  structs here for conciseness.}  \modelname{} is able to recover the
actual structure much of the time.
 At other times, \modelname{} also
produces some semantically reasonable but syntactically unacceptable
predictions, like \lt{char[32]} for \lt{class std::string}.

\subsection{RQ2: Comparison with Prior Work}
\label{sec:exp-osprey}

We further compare \modelname with
recent work on type recovery~\cite{zhang2021} and variable name recovery~\cite{lacomis2019dire}.

\mysec{Type Recovery}
While there is prior work on type recovery (see also \cref{sec:related}), none
of the existing approaches, TIE~\cite{Lee2011}, Howard~\cite{howard2011},
Retypd~\cite{Noonan2016}, TypeMiner~\cite{maier2019typeminer} and
\osprey~\cite{zhang2021}, are publicly available.
We are grateful to \citet{zhang2021}, the authors of \osprey,
for kindly sharing their evaluation material so we could compare results.

\osprey
is a recently proposed probabilistic technique for variable and
structure recovery that outperforms existing work including
Howard~\cite{howard2011},
Angr~\cite{shoshitaishvili2016sok}, Hex-Rays~\cite{hexrays}
and Ghidra~\cite{zhang2021}.
The \osprey authors provided us with the GNU coreutils\footnote{\url{https://www.gnu.org/software/coreutils/}} executables
they used in their evaluation, which were
compiled with \tt{-O0} to disable optimization.
We ran \modelname on these executables, but only evaluated on stack
and heap variables, since \osprey does not recover register variables. This benchmark consists of 101 binaries and 17,089
variables.
We also define two subsets of the dataset:
\begin{description}[itemsep=0pt,topsep=1pt]
\item[\textit{Visited}] A subset of 13,020 variables that
  are covered by BDA~\cite{zhang2019bda}, a binary abstract interpretation tool that \osprey relies on.
  \osprey is expected to perform better on these covered functions than uncovered functions,
  which we also report as \textit{Non-Visited}.\footnote{
    A majority of uncovered functions are unreachable from the entry point of the binary,
    and others are indirect call targets which BDA fails to analyze.
  }
  However, \modelname is not subject to this limitation.
\item[\textit{Struct}] A subset of 3,061 variables related to structure types.
  Following \osprey, we include structs allocated on the stack, pointers to
  structs on the heap, and arrays of structs. These variables do \emph{not} have
  to be in the Visited subset.
\end{description}

Because \modelname{} can predict up to 48,888 different types, each including the full syntactic and semantic information,
we convert its predictions in a post-hoc manner to make it comparable with \osprey.%
\footnote{
Specifically, we discard type names and field names.
For example, \lt{bool} and \lt{char} are both converted to \lt{Primitive_1},
which stands for a primitive type occupying 1 byte of memory,
\lt{const char *} and \lt{char *} are converted to \lt{Pointer<Primitive_1>},
and \lt{struct ImVec2 \{float x; float y;\}} converted to \lt{Struct<Primitive_4, Primitive_4>}.}

\cref{tab:osprey} compares the accuracies of both systems.
On the overall coreutils benchmark, \modelname{} slightly outperforms \osprey (76.8\% vs 71.6\%).
\osprey outperforms \modelname on the Visited subset, but as expected, performs worse on the Non-Visited functions.
Meanwhile, \modelname{} is more consistent on Visited and Non-Visited.
When only looking at structure types, \osprey outperforms \modelname (26.6\% vs 15.7\%).

However, this comparison puts \modelname at a disadvantage, since \osprey was
designed for this task of recovering syntactic types, while \modelname was
trained to recover variable and type/field names, and much of this information
is thrown out for this evaluation.
To address this, we trained a new model, \modelname{}$_{Light}$, on
\dataname, but tailored the training to \osprey's simplified task.  The accuracy
of this model is also reported in \cref{tab:osprey}.  As
expected, the \modelname{}$_{Light}$ model outperforms the
off-the-shelf \modelname model, since it is trained specifically for
this task.  \modelname{}$_{Light}$ greatly improves prediction accuracy
on the Struct subset, and even outperforms \osprey.

\begin{figure}
  \centering
  \includegraphics[width=0.75\linewidth]{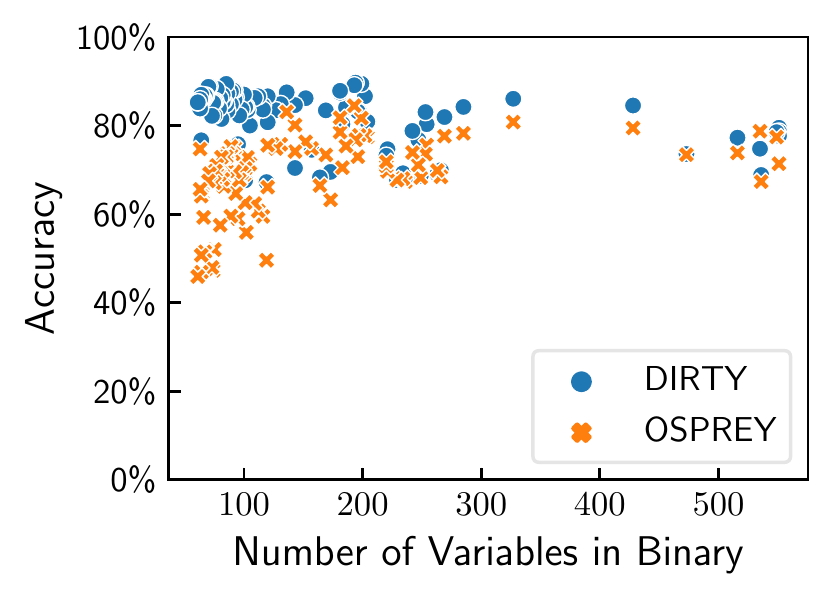}
  \caption{Accuracy of \modelname and \osprey on 101 individual programs in the coreutils benchmark with different number of variables.
  The two methods are competitive on large binaries,
  while \modelname{} performs much better on small binaries.}
  \label{fig:coreutils_breakdown}
\end{figure}

To further get a fine-grained comparison with \osprey,
we calculate accuracy on 101 coreutils binaries individually,
and show the prediction accuracies of \modelname and \osprey
with respect to the number of variables in the programs in \cref{fig:coreutils_breakdown}.

We observe that \modelname is competitive compared with \osprey.
Interestingly, while the results on large binaries are close,
\modelname performs better on small binaries.
This suggests our learning-based method trained on GitHub data might generalize better
on rare patterns compared to empirical methods that might have been developed based on
observations on a limited number of common and relatively larger programs.

In addition, \modelname{} is also much faster and scalable.
On average, \osprey takes around 10 minutes to analyze one binary in coreutils,
while it takes 75 seconds for \modelname{}$_{Light}$ to finish inference on the whole coreutils benchmark.

Overall, we believe both methods are valuable.
Since at this point \modelname{} is using Hex-Rays recovered data layout as input to its Data Layout Encoder,
we believe a promising future direction is to combine these two methods---using
\osprey's results as the input to \modelname's, and the combined approach can
potentially achieve even better results.

\begin{table}
  \centering
  \begin{tabular}{lrrrr}
      \toprule
       & \multicolumn{4}{c} {Coreutils} \\
      Model                 & All           & Visited       & Non-Visited   & Struct        \\
      \midrule
      \osprey               & 71.6          & \textbf{83.8} & 32.4          & 26.6          \\
      \modelname{}          & 76.8          & 79.1          & 69.6          & 15.7          \\
      \modelname{}$_{Light}$& \textbf{80.1} & 80.1          & \textbf{80.1} & \textbf{27.7} \\
      \bottomrule
  \end{tabular}
  \caption{Accuracy comparison on the Coreutils benchmark.}
  \label{tab:osprey}
\end{table}

\mysec{Name Recovery}
\label{sec:dire_comparison}
The Decompiled Identifier Renaming Engine (DIRE) is a state-of-the-art neural
approach for decompiled variable name recovery~\cite{lacomis2019dire}.
The DIRE model consists of both a lexical encoder and a structural encoder,
utilizing both tokenized decompiled code and the reconstructed abstract syntax
tree (AST).
In contrast, \modelname{}'s simpler encoder only uses the tokenized decompiled
code.

\begin{table}
    \centering
    \begin{tabular}{lrrrrrr}
        \toprule
         & \multicolumn{3}{c} {DIRE Dataset} & \multicolumn{3}{c} { \dataname{} Dataset}\\
        Model       & All            & FIT & FNIT   & All           & FIT & FNIT \\
        \midrule
        DIRE         & 72.8          & 84.1            & 33.5          & 57.5          & 75.6          & 31.8          \\
        \modelname{} & \textbf{81.4} & \textbf{92.6}   & \textbf{42.8} & \textbf{66.4} & \textbf{87.1} & \textbf{36.9} \\
        \bottomrule
    \end{tabular}
    \caption{Accuracy comparison of DIRE and \modelname{} on the DIRE and
      \dataname{} datasets. Accuracy is reported overall (All), when functions
      are in the training set (FIT), and when functions are not in the training set (FNIT). }
    \label{tab:dire}
\end{table}

The DIRE authors provide a public dataset for decompiled variable
renaming
compiled with \tt{-O0}.  To compare with DIRE, we train
\modelname{} on the DIRE dataset and also train DIRE on the
\dataname{} dataset.
Since DIRE is focused on variable renaming, and there is no type information
collected in their dataset, we cannot use the Data Layout Encoder for these
experiments. Instead, we only use our Code Encoder and Renaming
Decoder.
We report the accuracy of both systems in \cref{tab:dire}.
\modelname{} significantly outperforms DIRE in terms of overall
accuracy on both the DIRE dataset (81.4\% vs. 72.8\%), and on the
\dataname{} dataset (66.4\% vs. 57.5\%).
\modelname also generalizes better than DIRE: when functions are not
in the training set, \modelname{} outperforms DIRE on both the
DIRE (42.8\% vs. 33.5\%) and the \dataname{} datasets (36.9\%
vs. 31.8\%).

\modelname outperforms DIRE in spite of the fact that it only
leverages the decompiled code, whereas DIRE leverages both the
decompiled code \emph{and} the reconstructed AST from Hex-Rays.
Since the primary difference between \modelname without type prediction and DIRE
is that it uses Transformer as its encoder and decoder network,
we
attribute this improvement to the power of Transformers, which allow modeling
interactions between any pair of tokens, unrestricted to a sequential
or tree structure as in DIRE.

Also notable is how \modelname trains faster than DIRE.  We found that
\modelname surpassed DIRE in accuracy after training for 30 GPU hours, compared
to the 200 GPU hours required to train DIRE on the full \dataname dataset,
which we again attribute to the efficiency
of the Transformer architecture.

\subsection{RQ3: Ablation Study}

To understand how each component of \modelname contributes to its
overall performance, we perform an ablation study.

\begin{table}
    \centering
    \begin{tabular}{lrr}
        \toprule
         &  \multicolumn{2}{c}{Accuracy}                                             \\
        Model                  &  Overall      & Struct        \\
        \midrule
        \modelname{}$_{S}$     & 74.5          & 65.4          \\
        \modelname{}           & \textbf{75.8} & \textbf{68.6} \\
        \bottomrule
    \end{tabular}
    \caption{Effect of model size.
      The accuracy columns show the overall accuracy and
      the accuracy on struct types.
      }
    \label{tab:model_size}
\end{table}

\mysec{Model Size}
\label{sec:ablation_model_size}
Transformers have the merit of scaling easily to larger representational power
by stacking more layers, increasing the number of hidden units and attention heads
per layer~\cite{transformer,devlin2019bert}.
We compare \modelname{} to a modified, smaller version
\modelname{}$_{S}$.
\modelname{} contains 167M parameters, while \modelname{}$_{S}$ only 40M.
\cref{tab:hyper} contains details of the hyperparameter differences between the two models.

\cref{tab:model_size} shows
overall \modelname{} is 75.8\% accurate
vs. 74.5\% for \modelname{}$_{S}$'s.
This indicates increasing the model size has a positive effect on retyping
performance.
The gain from increased model capacity is notably larger when comparing
performance on structures.
This improvement suggests that complex types are more challenging and require a
model with larger representational capacity.
We are not able to train a larger model due to limits on computation power.

\begin{figure}[t]
    \centering
  \includegraphics[width=0.75\columnwidth]{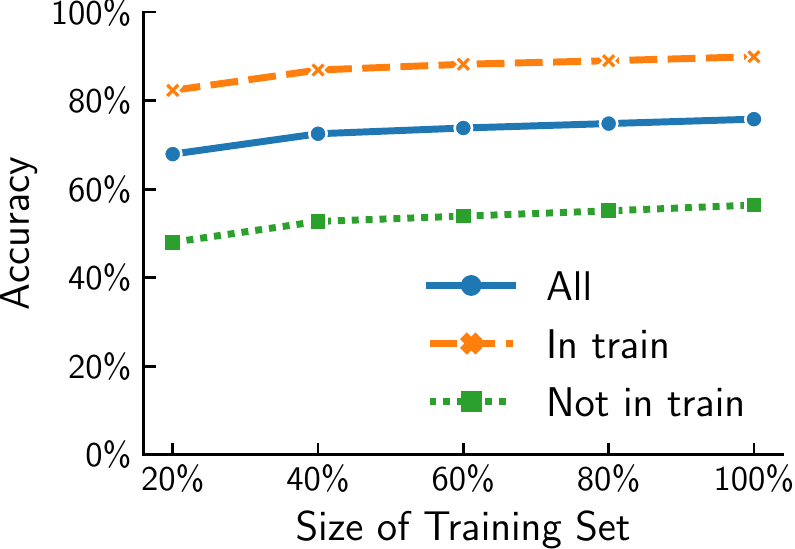}
  \label{fig:data_size_overall}
    \caption{Effect of training data size. With 100\% of the data, the
      accuracies of \emph{All}, \emph{In train}, and \emph{Not in train} are
      75.8\%, 89.9\%, and 56.4\% respectively. With 20\%, these drop to 67.9\%,
      82.3\%, and 48.0\% respectively.}
    \label{fig:data_size}
\end{figure}

\mysec{Dataset Size}
\label{sec:ablation_data}
We examine the impact of training data size on prediction accuracy.
As a data-driven approach, \modelname{} relies on a large-scale code dataset;
studying the impact of data size gives us insight into the amount of data to collect.
We trained \modelname{} on 20\%, 40\%, 60\%, 80\% and 100\%
portion of the full training partition
and report the results in
\cref{fig:data_size}.

\cref{fig:data_size} shows the change in accuracy with respect to the percentage of
training data.
Increasing the size of training data has a significant positive effect on the
accuracy.
Between 20\% and 100\% of the full size the accuracy increases from
67.9\% to 75.8\%, a relative gain of 11.6\%.

Notably, accuracy on \textit{Function not in training} has a relative gain of
17.5\% much larger
than on the \textit{Function in training} partition.
This is likely because the \textit{Function in training} partition contains
common library functions shared by programs both in the training and test
set, and even a smaller dataset will have programs that use these functions.
In contrast, the \textit{Function not in training} part is open-ended and
diverse.

It is also worth noting that the accuracy drops sharply when the training set
size is decreased from 40\% to 20\%, justifying the necessity for using a
large-scale dataset.

\mysec{Data Layout Encoder}
\label{sec:data_layout_encoder}
We explore the impact of the Data Layout encoder on
\modelname{}'s performance.
We experiment with a new model with no Data Layout encoder,
\modelname{}$_{NDL}$.

\begin{table}
    \centering
    \begin{tabular}{lrrr}
        \toprule     Model & Overall       & In train      & Not in train  \\
        \midrule
        \modelname{}$_{NDL}$     & 72.2          & 88.4          & 49.9          \\
        \modelname{}           & \textbf{75.8} & \textbf{89.9} & \textbf{56.4} \\
        \bottomrule
    \end{tabular}
    \caption{Effect of the Data Layout encoder on the accuracy of
      \modelname{}. Accuracy is reported for the model with (\modelname{}) and
      without (\modelname{}$_{NDL}$) the encoder.
      }
    \label{tab:mem}
\end{table}

\cref{tab:mem} shows the accuracy results overall and on the \textit{Function in training} and \textit{Function not in training} partitions.
The inclusion of the Data Layout encoder improves overall accuracy from
72.2\% to 75.8\%, indicating that the Data Layout encoder is effective.
The results are even more interesting when the results are broken into the two
partitions.
The relative gain on the \textit{Function in not training} partition is 13\%
(49.9\% to 56.4\%), compared to 1.7\% on the \textit{Function in training}
partition (88.8\% to 89.9\%). This suggests
the Data Layout encoder greatly improves \modelname{}'s generalization ability.

\begin{table*}
    \centering
    \small
    \begin{tabular}{lrlrlrlr}
        \toprule
        \multicolumn{4}{c}{\modelname{}} & \multicolumn{4}{c}{\modelname{}$_{NDL}$} \\
        \multicolumn{2}{c}{\lt{\_\_int64}} & \multicolumn{2}{c}{\lt{struct \_\_m128d}} & \multicolumn{2}{c}{\lt{\_\_int64}} & \multicolumn{2}{c}{\lt{struct \_\_m128d}} \\
        \midrule
        \lt{\_\_int64}    & 74.3\% & \lt{struct \_\_m128d}         & 78.7\% & \lt{\_\_int64}                & 67.0\% & \colorbox{gray!30}{\lt{double}}    & 33.1\%  \\
        \disappear    &  5.7\% & \disappear                & 15.4\% & \colorbox{gray!30}{\lt{int}}          &  6.3\% & \disappear             & 27.2\%  \\
        \lt{void *}       &  1.7\% & \colorbox{gray!30}{\lt{void}} &  2.9\% & \disappear                &  6.0\% & \colorbox{gray!30}{\lt{\_\_int64}} & 10.3\%  \\
        \lt{char *}       &  1.7\% & \lt{\_\_int128}               &  2.2\% & \colorbox{gray!30}{\lt{unsigned int}} &  1.5\% & \lt{struct \_\_m128d}      &  5.9\%  \\
        \lt{const char *} &  1.6\% & \colorbox{gray!30}{\lt{double}}       &  0.7\% & \lt{char *}                   &  1.2\% & \colorbox{gray!30}{\lt{int}}       &  3.7\%  \\
        \bottomrule
    \end{tabular}
    \caption{Comparative examples from \modelname{} with and without
      Data Layout encoder from the \textit{Function not in training}
      partition. Predictions inside a
      {\setlength{\fboxsep}{1pt}\colorbox{gray!30}{gray box}} have a
      different data layout than the ground truth type. \modelname{}
      effectively suppresses these, which helps guide the model to a
      correct prediction. The structure's full type is \lt{struct
        \_\_m128d \{double[2]} \lt{m128d\_f64;\}}.}
    \label{tab:preds_mem}
\end{table*}

\cref{tab:preds_mem} compares example predictions from \modelname{} and
\modelname{}$_{NDL}$ on the same types from the \textit{Function not in
  training} partition.
For the \lt{\_\_int64} example, the type predictions from \modelname{} mostly
have the correct size of 8 bytes.
\modelname{}$_{NDL}$, however, often incorrectly predicts \lt{int} and
\lt{unsigned int}.
This is understandable because in situations where the value doesn't exceed the
32-bit integer, \lt{\_\_int64} can be safely interchanged with \lt{int}, these
situations can be identified in some decompiled code.
However, apart from the correctness of the retyped program, accuracy to the
original binary, (\ie allocating 8 bytes instead of 4), is also important.
\modelname{} achieves this better than \modelname{}$_{NDL}$.

In the second example,
the \lt{struct \_\_m128d} type occupies 16 bytes, and has two members at offset
0 and 8.
\modelname{}$_{NDL}$ mainly mistakes this structure as a \lt{double}, which
might make sense semantically but is unacceptable syntactically.
With the Data Layout encoder, \modelname{} effectively reduces these errors.
This demonstrates this component achieves the soft masking effect on type
prediction as intended in \cref{sec:datalayout}.

\begin{table}
    \centering
    \begin{tabular}{lrrrr}
      \toprule
      & \multicolumn{2}{c}{Retyping} & \multicolumn{2}{c}{Renaming} \\
      Model & Overall & \checkmark Name        & Overall        & \checkmark Type        \\
      \midrule
      Retyping        & \textbf{75.8} & 90.6          & -             & -             \\
      Renaming        & -             & -             & \textbf{66.4} & 82.6          \\
      Multi-Task        & 74.9          & \textbf{92.3} & 65.1          & \textbf{84.6} \\
      \bottomrule
    \end{tabular}
    \caption{Performance comparison of the Retyping-only, Renaming-only, and
      Multi-Task decoders. Overall performance is shown, in addition to
      performance on retyping when the name is correct (\checkmark Name) and
      performance on renaming when the type is correct (\checkmark Type).}
    \label{tab:mt}
\end{table}

\mysec{Multi-Task Decoder}
\label{sec:exp_multitask}
In this section we study the effectiveness of the Multi-Task decoder when
compared to decoders designed for only retyping or only renaming.
Inspecting the accuracy numbers reported in \cref{tab:mt},
the Multi-Task decoder has similar, but slightly lower overall accuracy on both tasks
as the two specialized models (-0.8\% for retyping and -1.3\% for renaming).
One possible reason is that the Multi-Task model has twice the length of
decoding lengths than a specialized model, which makes greedy decoding harder.

Despite the small decrease in performance, the unified model has advantages.
These are illustrated in the \checkmark Name and \checkmark Type columns of
\cref{tab:mt}.
\checkmark Name and \checkmark Type stand for the subsets of the full dataset
where the Multi-Task decoder makes correct renaming predictions and correct
retyping predictions, and we evaluate the retyping and renaming performance on
them, respectively.%
\footnote{The probability of success on the other task also increases by chance,
  because success on one task implies it is easier than average.  We have
  eliminated this influence by, \eg comparing 92.3 to 90.6, instead of
  74.9.}
The Multi-Task decoder outperforms the specialized models by 1.9\%
and 2.4\% relatively on these metrics, in spite of the longer decoding
length.
This means the type and name predictions from the Multi-Task decoder are more
consistent with each other than from specialized models.
In other words, making a correct prediction on one task increases the
probability of success on the other task.

In practice, this offers additional flexibility and opens the opportunity for
more applications.
For example, consider a cooperative setting where a human decompilation expert
uses \modelname{} as an analysis tool.
The human expert may be unsatisfied with the model's top prediction and want to
switch to another one in the top-k candidates list.
With a Multi-Task decoder, the model adjusts the name prediction
for that variable, which is impossible with the specialized decoders.

\begin{table}
  \centering
  \begin{tabular}{lrrrr}
      \toprule
       & \multicolumn{4}{c} {GNU coreutils} \\
      Model       & \tt{-O0}  & \tt{-O1} & \tt{-O2} & \tt{-O3} \\
      \midrule
      \modelname{}& 48.20 & 46.01 & 46.04 & 46.00 \\
      \bottomrule
  \end{tabular}
  \caption{Accuracy comparison of \modelname{} on the GNU coreutils benchmark compiled with \tt{-O0}, \tt{-O1}, \tt{-O2}, and \tt{-O3} optimization levels.}
  \label{tab:optimization}
\end{table}

\subsection{RQ4: Compiler Optimization Levels}
\label{sec:optimization_levels}
We study the impact of compiler optimizations on \modelname{}'s accuracy.
In keeping with the spirit of the \osprey evaluation on coreutils compiled with \tt{-O3},
we choose coreutils as our evaluation dataset.
However, since we did not have access to the original dataset used by \osprey except \tt{-O0},
we recompiled GNU coreutils~3.2 ourselves using optimization levels \tt{-O0}, \tt{-O1}, \tt{-O2}, and \tt{-O3}.
\cref{tab:optimization} shows how accurately \modelname is able
to recover the full type (including type and field names) informaition at each optimization level.
As expected, \modelname does best at \tt{-O0}, since \modelname is
trained on \tt{-O0} code and we believe \tt{-O0} code to be
simpler.
Going from \tt{-O0} to \tt{-O1}, \modelname's accuracy drops from 48.2\% to
46.0\%.
However, there is little difference in performance between \tt{-O1}, \tt{-O2},
and \tt{-O3}.  This suggests that \modelname does slightly better on the
optimization level of code it was trained on, but that the effect of
optimizations is small.
We believe this is because Hex-Rays recognizes and will ``undo'' some
optimizations so that the decompiled code will be very similar.
For example, unoptimized code will often reference stack variables
using a frame pointer, but optimized code will reference such
variables using the stack pointer, or even maintain them in a
register.  But both implementations will look similar in the
decompiled code, since the mechanism used to reference the variable is
not important at the C level.
Since \modelname operates on the decompiled code, the decompiler
effectively insulates \modelname from these optimizations.

\subsection{Illustration}
\begin{figure}[t]
    \renewcommand\lt[1]{{\lstinline[style=cstyle3]!#1!}}
    \begin{lrbox}{\verbsavebox}
        \begin{lstlisting}[style=cstyle3]
int find_unused_picture(int a1, int a2, int a3) {
    int i, j, v1;
    if (a3) {
        for (i = <Num>;; ++i) {
        if (i > <Num>)
            goto LABEL_13;
        if (!*(*(<Num> * i + a2) + <Num>))
            break;
        }
        v1 = i;
    } else {
        for (j = <Num>;; ++j) {
        if (j > <Num>) {
        LABEL_13:
            av_log(a1, <Num>, <Str>);
            abort();
        }
        if (pic_is_unused(<Num> * j + a2))
            break;
        }
        v1 = j;
    }
    return v1;
}
        \end{lstlisting}
    \end{lrbox}
    \subcaptionbox*{}[\linewidth]{\usebox{\verbsavebox}}

    \begin{tabular}{lll}
        \toprule
        ID                       & Developer                     & \modelname{}              \\
        \midrule
        \textcolor{red}{\lt{a1}} & \lt{AVCodecContext\_0 *avctx} & \lt{MpegEncContext\_0 *s} \\
        \textcolor{red}{\lt{a2}} & \lt{Picture\_0 *picture}      & \lt{Picture\_0 *pic}      \\
        \textcolor{red}{\lt{a3}} & \lt{int shared}               & \lt{int shared}           \\
        \textcolor{red}{\lt{v1}} & \lt{int result}               & \lt{int result}           \\
        \bottomrule
    \end{tabular}
    \caption{Simplified Hex-Rays output. \lt{<Num>} and
        \lt{<Str>} are placeholder tokens for constant numbers and strings
        respectively. The table summarizes the original developer names and types
        along with the names and types predicted by \modelname{}.}
    \label{tab:case}
\end{figure}

To gain more qualitative insights into \modelname's predictions, consider the example
in \cref{tab:case}. 
The code shown is the Hex-Rays output, cleaned for presentation.
Here, we would like to rename and retype the arguments \lt{a1}, \lt{a2}, and
\lt{a3}, in addition to the variable \lt{v1}.
The table in \cref{tab:case} shows the developer's chosen types and names
together with \modelname{}'s suggestions.
\modelname{} suggests the same types and names as the developer for \lt{a3} and
\lt{v1}, and the same type but a different name for \lt{a2}.
Although the names disagree for \lt{a2}, we note that \lt{pic} is an
abbreviation for \lt{picture}, so the disagreement is minor.
We also observe that \lt{Picture\_0 *}, the type of \lt{a2} itself carries
a lot of semantic information; even if \modelname{} was unable to suggest
a meaningful name, \lt{Picture\_0 *a2} is still helpful for reverse
engineering.

The developer and \modelname{} disagree on both the name and the type of
\lt{a1}.
In this case, the name chosen by \modelname{} (\lt{s}) would probably not be
considered a very useful improvement over \lt{a1}.
However, the type suggested by \modelname{} (\lt{MpegEncContext\_0 *}) could
still be quite useful to a reverse engineer, even if it is inaccurate.
It suggests that this argument is a ``context'', and hints that this
function is used for video.

\section{Related Work}
\label{sec:related}

Other projects related to type recovery for decompilation are
REWARDS~\cite{lin2010automatic}, TIE~\cite{Lee2011}, Retypd~\cite{Noonan2016},
and \osprey~\cite{zhang2021}.  Unlike our approach, they use program
analyses to compute constraints on types.  Additionally, they are
either limited to only predicting the syntactic type (TIE, Retypd,
\osprey), or only predicting one of a small set of hand-written types
(150 for REWARDS).
In comparison, \modelname{} automatically
generates a database of types by observing real-world code.

Other projects use machine learning to predict types, but target different
languages than \modelname{}.
\textsc{DeepTyper}~\cite{Hellendoorn2018} learns type inference for JavaScript and
\textsc{OptTyper}~\cite{pandi2020opttyper}, \textsc{LambdaNet}~\cite{wei2019lambdanet},
$\text{R-GNN}_{\text{NS-CTX}}$~\cite{ye2020advanced} target TypeScript.
Training a machine learning algorithm for the task of typing dynamic languages like
these is a slightly easier task: generating a parallel corpus is simple, since
the types can simply be removed without changing the semantics.
The \dataname{} dataset is fundamentally different: including debug information
often changes the layout of the code as the decompiler adds structures and
syntax for accessing them.

To the best of our knowledge, the most directly-related work to \modelname{} is
TypeMiner~\cite{maier2019typeminer}.
TypeMiner is a pioneering work, providing the proof-of-concept for recovering
types from C binaries.
However, they use much simpler machine learning algorithms and their dataset only
consists of 23,482 variables and 17 primitive types.
Escalada~et~al.~\cite{escalada2021improving} has provided similar insights.
They adopt simple classification algorithms to predict function return types in
C, but they only consider from only 10 different (syntactic) types and their
dataset is limited to 2,339 functions from real programs and 18,000 synthetic
functions.

Two other projects targeting the improvement of decompiler output using neural
models are DIRE~\cite{lacomis2019dire}, which predicts variable names,
DIRECT~\cite{DIRECT}, which extends DIRE using transformer-based models,
and Nero~\cite{david2020neural}, which generates procedure names.  Other approaches work directly on assembly~\cite{Katz2018, katz2019towards,
  fu2019coda}, and learn code structure generation instead of aiming to recover
developer-specified variable types or names.  Similarly,
\textsc{Debin}~\cite{He2018} and CATI~\cite{Chen2020} use machine learning to
respectively predict debug information and types directly from stripped
binaries without a decompiler.

\section{Discussion}
\label{sec:future}

In this paper we presented \modelname{}, a novel deep learning-based technique for predicting
variable types and names in decompiled code.
Still, \modelname is limited in several ways that provide key opportunities for future
improvements.

\mysec{Alternative Decompilers to Hex-Rays} We implement \modelname on top of
the Hex-Rays decompiler because of its positive reputation and the programmatic
access it affords to decompiler internals. However, \modelname is not
fundamentally specific to Hex-Rays, and the technique should conceptually work
with any decompiler that names variables using DWARF debug symbols. Note that,
due to its recent popularity and promise, we attempted to evaluate our
techniques using the newer, open-source Ghidra decompiler. Unfortunately, it is
currently infeasible, because Ghidra routinely failed to accurately name stack
variables based on DWARF. This appears to be a combination of specific issues\footnote{\href{}{https://github.com/NationalSecurityAgency/ghidra/issues/2322}} and
the general design of the decompiler. Ghidra's decompiler consists of many
\emph{passes} which modify and augment the current decompilation. Some of these
passes combine variables, but in doing so may combine a DWARF-named variable
with others. Since the combined variable no longer corresponds directly with the
DWARF variable information, Ghidra discards the name.  We are optimistic,
however, that when the above-mentioned issues are addressed, Ghidra may again be
a reasonable target for our approach.

\mysec{Generalizing to Unseen Types}
A limitation of \modelname's current decoder is that it can only predict types
seen during training. Fortunately, there appears, empirically, to be sufficient
redundancy across large corpora that \modelname is still frequently able to
successfully recover structural types. This lends credence to the hypothesis
that code is \emph{natural}, an observation that has been explored in several
domains~\cite{Hindle2012,Devanbu2015}. It moreover appears that data layout is of particular
importance here: layout information recovered from the decompiler impose
key constraints on the overall prediction problem. Indeed, our results in
\cref{sec:data_layout_encoder} corroborate the intuition that the Data
Layout Encoder is especially important for succeeding on previously unseen code.

We envision meaningful future opportunities to more directly expand \modelname's
capabilities to predict unseen structures.  This
problem is analogous machine translation models that must deal with
rare or compound words (e.g., xenophobia) that are not present in their
dictionary.
Byte~Pair~Encoding~\cite{sennrich2015neural} (BPE) is the most frequently used
technique to tackle this problem in the natural language domain.
It automatically splits words into multiple tokens that are present in the dictionary (e.g., xeno and \#\#phobia).
(The \#\# indicates the word is still part of the current word, instead of a new word next to it.)
This technique greatly increases the number of words a model can handle
despite a limited dictionary size, and enables the composition of new words that
were not seen during training.
This suggests that we can similarly extend \modelname's decoder to predict previously unseen types by
decomposing structure types into multiple pieces with BPE.
  For example, a structure type \lt{struct timeval \{time\_t tv\_sec;} \lt{suseconds\_t tv\_usec;\}} is split into
  four separate tokens, which are 1) \lt{struct timeval}, 2) \lt{time\_t tv\_vec;}, 3) \lt{suseconds\_t tv\_usec;}, and 4) \lt{<end\_of\_struct>}.

However, unfortunately, our preliminary experiments suggested that this hurts overall prediction accuracy.
It also significantly slows down prediction,
since it drastically increases the number of decoding steps.  It moreover
requires finer-grained accuracy metrics, like tree distance, to allow us to
measure and credit partially correct predictions.
Based on these observations,
we believe unseen structure types should be handled specially with a tailored
model, a problem we leave to future work.

\mysec{Supporting Non-C Languages}
A benefit of decompiling to C is that as a relatively
low-level language, it can express the behavior of executables beyond
those written in C.
Although we designed \modelname to be used with C programs and types,
\modelname can run on non-C programs, and will try to identify the C
type that best captures the way in which that variable is being used.
Thus, \modelname provides value to analysts seeking to understand
non-C programs, similar to how C decompilers such as Hex-Rays help
analysts to understand C++ programs.

However, many compiled programming languages have type systems far
richer than C's, and expressing these types in terms of C types may be
confusing.
For example, in C++, virtual functions are often implemented by
reading an address out of a \emph{virtual function
  table}~\cite{schwartz:2018:ooanalyzer,erinfolami:2020}.
Although techniques like \modelname can recognize such tables as
structs or arrays of code pointers, it does not  reveal the
connection to the higher-level C++ behavior of virtual functions.

Extending \modelname to support higher-level languages such as C++ is
an interesting open problem.
To some degree, as long as the decompiler is able to import the
higher-level type information from debug symbols into the decompiler
output, it should be possible to train \modelname to recognize non-C
types.  For instance, 6\% of the programs in \dataname are written in
C++, and our evaluation measures \modelname's ability to predict
common C++ class types such as \lt{std::string}.
But recovering higher level properties of these types, especially for
those never seen during training, is a challenging problem and is
likely to require language-specific
adaptations~\cite{schwartz:2018:ooanalyzer,erinfolami:2020}.

\mysec{Limited Input Length}
As common with Transformers,
we truncate the decompiled function if the length $n$ exceeds some upper limit $max\_seq\_length$,
which makes training more efficient.
In our experiments we set $max\_seq\_length$ to 512 for two reasons.
First, 512 is the default value for $max\_seq\_length$ in many Transformer models~\cite{transformer,devlin2019bert}.
Second, in \dataname{}, the average number of tokens in a function is 220.3,
and only 8.8\% of the functions have more than 512 tokens, \ie we exclude relatively few of the possible inputs encountered in the wild.
Still, if enough computational resources are available, we recommend
using efficient Transformer implementations such as Big~Bird~\cite{zaheer2020big} instead.
These can deal with much larger $max\_seq\_length$ and can be used out-of-the-box to
replace our implementation.

\section{Conclusion}

The decompiler is an important tool used for reverse engineering.
While decompilers attempt to reverse compilation by transforming binaries into
high-level languages, generating the same code originally written by the
developer is impossible.
Many of the useful abstractions provided by high-level languages such as loops,
typed variables, and comments, are irreversibly destroyed by compilation.
Decompilers are able to deterministically reconstruct some structural properties
of code, but comments, variable names, and custom variable types are technically
impossible to recover.

In this paper we address the problem of assigning decompiled variables
meaningful names and types by statistically modeling how developers write code.
We present DIRTY (DecompIled variable ReTYper), a novel technique for improving
the quality of decompiler output that automatically generates meaningful
variable names and types.  Empirical evaluation of DIRTY on a novel dataset of C
code mined from GitHub shows that DIRTY outperforms prior work approaches by a
sizable margin, recovering the original names written by developers 66.4\% of
the time and the original types 75.8\% of the time.

\newpage






\balance

\bibliographystyle{plainnatetal}
\bibliography{references}

\newpage

\appendix

\section{Experimental Setup}
\label{sec:appendix}

\paragraph{Hyperparameter Configurations}
Our detailed hyperparameters are shown in \cref{tab:hyper}.
We use a six-layer Transformer Encoder for the code encoder,
a three-layer Transformer Encoder for the data layout encoder,
and a six-layer Transformer Decoder for the type decoder.
We set the number of attention heads to $8$.
Input embedding dimensions and hidden sizes $d_{model}$ are set to $512$ for the code encoder, and $256$ for the data layout encoder.
Following prior work, we empirically set the size of the inner-layer of the position-wise feed-forward inner representation size $d_{ff}$ to four times the hidden size $d_{model}$~\cite{transformer}.
We use the $gelu$ activation function~\cite{hendrycks2016gaussian} rather than the standard $relu$, following BERT~\cite{devlin2019bert}.
During training, we set the batch size to 64 and the learning rate to $1 \times 10 ^ { - 4 }$.
We use the Adam optimizer~\cite{adam} and set $\beta _ { 1 } = 0.9 , \beta _ { 2 } = 0.999 \text { and } \epsilon = 1 \times 10 ^ { - 8 }$.
We apply gradient clipping by value within the range $\left[-1, 1\right]$.
We also apply a dropout rate of $0.1$ as regularization.
We train the model for 15 epochs.
At the inference time, we use beam search to predict the types for each function with a beam size of $5$.

\paragraph{Hardware Configuration}
We conducted all experiments on Linux servers equipped with two Intel
Xeon Gold 6148 processors, 192GB RAM and
8
NVIDIA
Volta V100 GPUs.  We expect that a similar machine could reproduce the
full training and testing stage of our main experiments in 120 GPU hours.

\paragraph{Software}
We implemented our models with PyTorch~\cite{pytorch} version 1.5.1 and Python 3.6.
We plan to release our dataset, code and pre-trained models at publication time.

\begin{table}[H]
  \centering
    \begin{tabular}{lrr}
        \toprule
        Hyperparameter              & \modelname{} & \modelname{}$_{S}$ \\
        \midrule
        Max Sequence Length         & 512          & 512                 \\
        Encoder/Decoder layers      & 6/6          & 3/3                 \\
        Hidden units per layer      & 512          & 256               \\
        Attention heads             & 8            & 4                 \\
        Layout encoder layers       & 3            & 3                 \\
        Layout encoder hidden units & 256          & 128               \\
        Batch size                  & 64           & 64                \\
        Training epochs             & 15           & 30                \\
        Learning rate                & $10^{-4}$    & $10^{-4}$        \\
        Adam $\epsilon$              & $10^{-8}$    & $10^{-8}$        \\
        Adam $\beta_1$              & 0.9          & 0.9               \\
        Adam $\beta_2$              & 0.999        & 0.999             \\
        Gradient clipping           & 1.0          & 1.0               \\
        Dropout rate                & 0.1          & 0.1               \\
        Number of parameters        & 167M         & 40M               \\
        \bottomrule
    \end{tabular}
    \caption{Summary of the hyperparameters of \modelname{} and the smaller
      \modelname{}$_{S}$.}
    \label{tab:hyper}
\end{table}

\begin{table}[H]
  \centering
  \begin{tabular}{lrr}
      \toprule
      Dataset                           & \modelname{} \\
      \midrule
      \#Binaries                        & 75,656       \\
      Unique \#functions (train)        & 718,765      \\
      Unique \#functions (valid)        & 139,473      \\
      Unique \#functions (test)         & 139,394      \\
      \% func body in train (valid)     & 64.6\%       \\
      \% func body in train (test)      & 65.5\%       \\
      Avg. \#code tokens                & 220.3        \\
      Median \#code tokens              & 86           \\
      Avg. \#identifiers per function   & 5.1          \\
      Median \#identifiers per function & 3            \\
      \bottomrule
  \end{tabular}
  \caption{Statistics of the \dataname{} datasets.}
  \label{tab:dataset}
\end{table}

\end{document}